# Statistics of genome architecture

V.R. Chechetkin[*]

*Theoretical Department of Division for Perspective Investigations, Troitsk Institute of Innovation and Thermonuclear Investigations (TRINITI), Troitsk-Moscow,*

*142190 Russia*

**Abstract**

The main statistical distributions applicable to the analysis of genome architecture and genome tracks are briefly discussed and critically assessed. Although the observed features in distributions of element lengths can be equally well fitted by the different statistical approximations, the interpretation of observed regularities may strongly depend on the chosen scheme. We discuss the possible evolution scenarios and describe the main characteristics obtained with different distributions. The expression for the assessment of levels in hierarchical chromatin folding is derived and the quantitative measure of genome architecture inhomogeneity is suggested. This theory provides the ground for the regular statistical study of genome architecture and genome tracks.

---

[*]*E-mail addresses:* chechet@biochip.ru and vladimir_chechet@mail.ru.

## 1. Introduction

Dynamic alteration of chromatin folding affects the gene regulation and the basic genetic processes such as DNA replication and recombination [1]. The mode of chromatin folding is termed genome architecture. The folding depends on the distribution of structural proteins over genome. The statistics of protein binding sites and, generally, the distribution of the other characteristics like transcription starts, length of protein-coding and non-coding regions, stretches of different nucleotide content, double-strand DNA breaks, contacts between chromosomes, etc. (termed commonly genome tracks) remains still the challenging problem from both experimental and theoretical point of view [2–4]. The terminal regions of tracks are much narrower than the mean length between consecutive termini and the number of terminal regions is several orders of magnitude less than the length of genome in base pairs (bp). Therefore, the set of sites implicated to genome architecture or genome tracks corresponds statistically to the sparse systems. In the main approximation the problem is reduced to the statistical analysis of length distribution between the nearest terminal sites (such fragments will be called below for definiteness as the elements of genome architecture or genome elements). In this Letter we briefly review and critically assess the main statistical distributions applicable to the analysis of genome architecture or genome tracks. The main original results presented in this Letter are: (i) application of De Finetti distribution to the study of genome tracks and genome architecture elements; (ii) the generalization of De Finetti distribution to the discrete case with termini of finite size for the element lengths; (iii) model of molecular evolution based on De Finetti statistics; (iv) equation for the assessment of levels of folding in genomes of different organisms; (v) quantitative measure of regularity for length distribution and of genome architecture complexity. This theory may be implemented to the development of a statistical tool for comparative genome-wide analysis of genome tracks and architecture elements.

## 2. De Finetti distribution

*2.1. Continuous De Finetti distribution*

Let the complete genome of length *M* be divided by *N* fragments of lengths, $L_1, ..., L_N$,

$$\sum_{k=1}^{N} L_k = M \qquad (1)$$

The resulting random division of genome by $N - 1$ points may be described in terms of De Finetti distribution [5]. The probability that the lengths of elements exceed the given thresholds is defined as

$$P(L'_1 \geq L_1; ...; L'_N \geq L_N) = \left(1 - L_1/M - ... - L_N/M\right)_+^{N-1} \qquad (2)$$

where $x_+ = x$, if $x > 0$ and $x_+ = 0$, if $x \leq 0$. The analytical derivation of probability (2) and the detailed calculations of various characteristics related to De Finetti distribution can be found in Ref. [6]. Here we reproduce only the main results needed for the subsequent consideration. The probability (2) can be conveniently presented using the lengths normalized to the mean,

$$l_k = L_k / <L>; <L> = M/N \qquad (3)$$

$$\sum_{k=1}^{N} l_k = N \qquad (4)$$

$$P(l'_1 \geq l_1; ...; l'_N \geq l_N) = \left(1 - l_1/N - ... - l_N/N\right)_+^{N-1} \qquad (5)$$

The corresponding one-element probability and probability density (henceforth the capital letter *P* will denote the probability, whereas the small letter *p* denotes the probability density) are given by

$$P(l'_1 \geq l; l'_2 \geq 0; ...; l'_N \geq 0) = (1 - l/N)_+^{N-1} \approx e^{-l};$$
$$p(l) = (1 - l/N)_+^{N-2}(N-1)/N \qquad (6)$$

At the large *N*, one-element De Finetti probability can be approximated by the exponential distribution depending only on the normalized length. The moments of normalized length are equal to

$$<l^k> = N^k \frac{k!(N-1)!}{(k+N-1)!} \qquad (7)$$

whereas the distribution of sum, $l_1 + ... + l_k = S_k$, in the exponential approximation is determined by

$$P(S_k \geq S) \approx \sum_{n=0}^{k-1} \frac{S^n}{n!} e^{-S};$$
$$p_k(S) = \frac{S^{k-1}}{(k-1)!} e^{-S} \qquad (8)$$

*2.2. Discrete De Finetti distribution*

In genetics, the various characteristics are defined in the discrete sites of genomic sequences, $m = 1, ..., M$. The terminal "points" of genome division correspond often to the binding regions of structural proteins having the finite length *a*. The different binding regions of length *a* are assumed to be non-overlapping. The corresponding discrete generalization of one-element probability is then given by

$$P(m' > m) = \frac{(M-m)(M-(m+a))...(M-(m+a(N-2)))_+}{(M-a)(M-2a)...(M-a(N-1))} \theta(m-a) \qquad (9)$$

where θ(x) is Heaviside step function. The derivation of probability (9) is based on the excluded volume effects. The first binding region of length *a* can occupy the relative number of sites $(M - m)/(M - a)$, the second non-overlapping region can occupy the relative number of sites $(M - (m + a))/(M - 2a)$, etc. The discrete moments are calculated as

$$<m^k> = \sum_{m=a}^{M-a(N-2)} m^k \left( P(m'>m) - P(m'>m+1) \right) \qquad (10)$$

If genome is covered by *W* non-overlapping windows and the characteristics over sites are replaced by the coarse-grained characteristics over windows, the distribution of windows with properties exceeding given threshold will again be determined by probability similar to (9), where *m* means the ordinal number of a window, $a = 1$, and $W \to M$.

The minimum of *k* distances between the centers of non-overlapping binding regions of width *a* will be determined by probability,

$$P_{\min|k}(m'>m) = \frac{(M-km)(M-(km+a))...(M-(km+a(N-2)))_+}{(M-ka)(M-(k+1)a)...(M-a(k+N-2))} \theta(m-a) \qquad (11)$$

If the binding regions of two types may overlap with each other, while the regions of one type remain non-overlapping, the probability that the distance between centers of binding regions of different types exceeds given threshold is determined as

$$P_{AB}(m'>m) = \frac{(M-m)(M-(m+a))...(M-(m+a(N_A-2)))_+}{M(M-a)...(M-a(N_A-2))}$$
$$\frac{(M-m)(M-(m+b))...(M-(m+b(N_B-2)))_+}{M(M-b)...(M-b(N_B-2))} \qquad (12)$$

This probability can serve for the assessment of overlapping and neighboring between binding regions of different types.

*2.3. Evolution of genome architecture*

The asymptotic exponential approximation of De Finetti distribution (6) possesses the remarkable property of statistical robustness, the random removal or addition of boundary points for genome elements affects only the mean length of elements and retains the form of exponential distribution. The change of form of probability distribution may be related with evolution inhomogeneous over genome. In the simplified evolution scenario the genome elements may be approximately divided by the group of relatively conservative elements and the group of rapidly evolving elements. Such scenario does not contradict the actual molecular evolution of genome architecture [7, 8]. Let in the group of rapidly evolving elements be permissible: (i) the fragmentation of a part of elements into the shorter ones and (ii) the aggregation of neighboring elements into the longer units. The shorter elements produced during fragmentation can be reshuffled over genome during subsequent evolution. If the number of fragments obeys Poisson statistics

$$P_{f,n} = \frac{\bar{n}_f^{n-1}}{(n-1)!} e^{-\bar{n}_f} \qquad (13)$$

the probability of finding a fragment with length exceeding given threshold is

$$P_f(l' \geq l) = \sum_{n=1}^{\infty} P_{f,n} \int dl'\, e^{-l'} (1 - l/l')_+^{n-1} = \int_l^{\infty} dl'\, e^{-l' - \bar{n}_f\, l/l'} \qquad (14)$$

where all lengths are measured relative to the mean length of elements before fragmentation. The mean fragment length corresponding to the distribution (14) is $<l>_f \approx 1/\bar{n}_f$ (here $\bar{n}_f$ is the mean number of the shorter elements produced during fragmentation of an initial element). In the case of Poisson aggregation of neighboring elements

$$P_{a,n} = \frac{(\bar{n}_a - 1)^{n-1}}{(n-1)!} e^{-(\bar{n}_a - 1)} \tag{15}$$

the corresponding probability density for aggregated lengths will be defined as (cf. the lower Eq. (8))

$$p_a(l) = \sum_{n=1}^{\infty} P_{a,n} \frac{l^{n-1}}{(n-1)!} e^{-l} = e^{-(\bar{n}_a - 1)} e^{-l} I_0\left(2((\bar{n}_a - 1)l)^{1/2}\right) \tag{16}$$

Here $I_0(x)$ is Bessel function of the zero order depending on imaginary argument. Its asymptotic dependence at the large arguments is

$$I_0(x) \approx \frac{1}{\sqrt{2\pi x}} e^x \tag{17}$$

The mean aggregated length corresponding to density (16) is $<l>_a = \bar{n}_a$ (here $\bar{n}_a$ is the mean number of the neighboring elements merged during aggregation). The lengths are again measured relative to the mean length of genome elements before aggregation. The resulting probability corresponds to the mixture of initial exponential distribution and the additives related to the fragmentation and aggregation processes

$$P(l' \geq l) = (1 - f_a - f_f) e^{-l} + f_f P_f(l' \geq l) + f_a P_a(l' \geq l) \tag{18}$$

It depends on five parameters: the mean length of architecture elements before evolutionary modifications $<L>$; fraction of fragmented elements $f_f$; fraction of aggregated elements $f_a$; two parameters $\bar{n}_f$ and $\bar{n}_a$ characterizing the fragmentation and aggregation processes.

## 3. Aggregation of elements and gamma-distribution

If the probability of aggregation is sharply peaked, the resulting probability density for aggregated elements can be approximated by the gamma-distribution

$$p(L) = \frac{L^{\alpha-1}}{\Gamma(\alpha)\beta^\alpha} \exp(-L/\beta) \qquad (19)$$

with two parameters $\alpha > 1$ and $\beta$. The probability density (19) describes the distribution of the resulting lengths of aggregated elements (cf. Eq. (8)). Index $\alpha$ indicates the mean number of aggregated elements and $\beta$ corresponds to their mean length before aggregation. The corresponding moments for the lengths of aggregated elements are

$$<L^k> = \beta^k \Gamma(\alpha+k)/\Gamma(\alpha) = \beta^k(\alpha+k-1)...\alpha \qquad (20)$$

The moments for normalized length $l = L/<L>$ depend only on $\alpha$, $<l^k> = (\alpha+k-1)...\alpha/\alpha^k$, and tend to unity at the large $\alpha$. The particular examples of application of gamma-distribution to the distribution of gene lengths can be found in [9–11].

## 4. Consecutive fragmentation and log-normal distribution

Log-normal distribution describes the process of random consecutive fragmentation. The typical examples are related to the formation of grains within minerals, the cascaded decay of vortices in fully-developed turbulence, the loss of particle energy during consecutive collisions, etc. In genetics, log-normal-like distributions appeared in transcription kinetics (RNA copying of DNA fragments coding for proteins) [12, 13]. The resulting transcription dynamics may be presented as the re-distribution of a pool of RNA monomers in a cell over poly-RNA strands synthesized during transcription of different coding DNA fragments. This yields log-normal-like transient dynamics and asymptotic power-like distributions for transcription intensity.

Consider the simplest example of consecutive random fragmentation [5]. Let the complete genome of length $M$ be consecutively fragmented by the architecture elements. After $n$ fragmentations, the resulting length of an element $L$ will be

$$L = f_1 f_2 ... f_n M \qquad (21)$$

where the random fractions $f_i$ are independent and uniformly distributed within the interval (0, 1). As the probability for the variable $y = -\ln f$ is exponential, $P(y' > y) = e^{-y}$, the resulting probability density for $\ln(M/L)$ is given by analog of Eq. (8) (up to the factor related to the replacement of variables),

$$p_n(L) = \frac{(\ln(M/L))^{n-1}}{(n-1)!} \qquad (22)$$

which yields the moments

$$<L^k> = M^k/(k+1)^n \qquad (23)$$

At the large $n$ the probability density (22) can be approximated as

$$p_n(L) \approx \frac{1}{L} \exp\left(-\frac{(\ln(M/L)-(n-1))^2}{2\ln(M/L)}\right) \approx \frac{1}{L} \exp\left(-\frac{(\ln(M/L)-(n-1))^2}{2(n-1)}\right) \qquad (24)$$

Kolmogorov [14] considered the generalized fragmentation processes and obtained the conditions of convergence to the log-normal distribution of general form

$$p(L) = \frac{1}{L\sigma\sqrt{2\pi}} \exp\left(-\frac{(\ln(L)-\mu)^2}{2\sigma^2}\right) \qquad (25)$$

governed by two parameters $\mu$ and $\sigma$. The density (25) provides the moments

$$<L^k> = \exp(k\mu + k^2\sigma^2/2) \qquad (26)$$

The corresponding probability density and the moments for normalized length $l = L/<L>$ are given by

$$p(l) = \frac{1}{l\sigma\sqrt{2\pi}} \exp\left(-\frac{(\ln(l)+\sigma^2/2)^2}{2\sigma^2}\right) \qquad (27)$$

$$<l^k> = \exp(k(k-1)\sigma^2/2) \qquad (28)$$

The dynamical range of $\ln(l)$ is limited by $\ln(l)_{min}$ and $\ln(l)_{max}$. The moments at the large $k$ tend to converge at the boundary of dynamical range, whereas the dependence in exponent (28) becomes linear in this limit [15].

## 5. Levels of chromatin folding

The value of consecutive fragmentations $(n-1)$ in Eq. (24) can be related to the hierarchical folding of chromatin DNA. The lowest level of folding in genomes of various organisms (except some viruses with RNA and single-stranded DNA genomes) starts with wrapping DNA around special histone-like proteins. Therefore, the shortest length in the hierarchical folding corresponds to the persistence length of double-stranded DNA, $L_p$ (~150 bp). The total number of chromatin folding levels grows approximately logarithmically with the length of genome $M$,

$$N_f \propto \log_{10}(M/L_p) \qquad (29)$$

Here $M$ should be understood as the mean length of genomes with given levels of chromatin folding (i.e. the dependence in Eq. (29) should be understood as a trend). The decimal logarithm in Eq. (29) is chosen because the hierarchical folding obeys approximately the rule: the higher level of folding contains about ten units of the preceding lower level (the actual variations in unit number are ranged within 5–20). Table 1 illustrates the modes of chromatin folding in different organisms (see also Supplemental materials). The relevant values of logarithms are $\log_{10}(M/L_p) \approx$ 1.5; 4.5 and 7.3 and turn out rather close to the corresponding number of folding levels. The characteristic lengths implicated to chromatin folding can be displayed in the related genomic DNA sequences by Fourier analysis (reviewed in [19]). For example, the symmetry of the fifth order in icosahedral capsid packing of viral genomes is associated with periods $M/5$ found in the viral genomic nucleotide sequences [20]. The cascade of characteristic lengths from the large scales to the shorter scales in the hierarchical chromatin folding may generate the log-normal-like distribution. The systematization of folding levels is yet absent.

## 6. Extreme value statistics

The characteristic maximum and minimum lengths of genome architecture elements or genome tracks depend on the total number of elements $N$. The corresponding normalized lengths $l = L/<L>$ can be estimated by the relationships

$$NP(l' > l_{max}) \approx 1; \; NP(l' < l_{min}) \approx 1 \tag{30}$$

This yields the following characteristic extreme values for the exponential, gamma- and log-normal distributions, respectively,

$$l_{max} \approx \ln N; \; l_{min} \approx 1/N \tag{31}$$

$$l_{max} \cong \ln N / \alpha + \ln \ln N; \; l_{min} \cong (\Gamma(\alpha)/N)^{1/\alpha} / \alpha \tag{32}$$

$$\ln l_{max/min} \cong \pm \sigma (\ln N)^{1/2} \tag{33}$$

The higher the number of elements $N$, the broader the dynamical range between $l_{min}$ and $l_{max}$. The dependence of the extreme lengths on $N$ should be taken into account at the assessment of dependence for the characteristic length of genome architecture elements or genome tracks on the complete genome size during molecular evolution.

For example, the mean gene codes for about 300 amino acids ($<L>$ = 900 bp). *E. coli* genome contains about 3,000 genes (ln $N$ = 8.0); the longest gene codes for 1,538 amino acids ($L$ = 4,614 bp) and is distinctly shorter than the expected value $<L>$ln $N$. Such behavior agrees with the bias against long genes in eubacteria because of the demanding costs of time and resources for protein production related to long genes [21]. Human genome contains about 30,000 genes (ln $N$ = 10.3); the longest gene codes for about 3,500 amino acids ($L$ = 10,500 bp) and is about the expected maximum statistical value. The example illustrates how extreme value statistics helps in discerning between natural selection load and unbiased random molecular evolution.

The minimum length may be short enough to be protein-coding region. This indicates that generally the length distribution for genes should be studied for the set uniting both coding and non-coding regions. The study of length distribution shed light on the pathways of molecular evolution and selection of elements.

**7. Inhomogeneity of genome architecture**

The variations in lengths of genome architecture elements characterize approximately the inhomogeneity of genome architecture. It can be proved that the structural entropy

$$S_{structural} = \sum_{i=1}^{N} l_i \ln l_i \tag{34}$$

at the restriction (4) attains its minimum equal to zero when all normalized lengths of genome architecture elements are equal to each other, $l_1 = l_2 = \ldots = l_N = 1$. This distribution corresponds to the most ordered (or "homogeneous") architecture. The higher the value of the structural entropy, the stronger the variations in characteristic lengths of genome architecture elements or the higher the inhomogeneity of architecture. The maximum entropy is attained when one of the lengths exceeds strongly all others. The mean value and dispersion of structural entropy for exponential distribution of normalized lengths are

$$<S_{structural}>_{exp} = (1-C)N = 0.422785...N; \quad <(\Delta S_{structural})^2>_{exp} = 0.289868...N \qquad (35)$$

where $C = 0.577215...$ is the Euler constant. The corresponding mean values for gamma- and log-normal distributions

$$<S_{structural}>_{gamma} = [\psi(\alpha+1) - \ln\alpha]N; \quad <S_{structural}>_{log-normal} = \sigma^2 N/2 \qquad (36)$$

where $\psi(x) = \Gamma'(x)/\Gamma(x)$, prove to be lower than that for the exponential distribution. This statement is valid for the gamma-distribution if $\alpha > 1$ and for the log-normal distribution with the first and second moments coincident with their counterparts for the exponential distribution. The comparison of architecture complexity for the genomes with different numbers of elements can be performed with the structural entropy per element,

$$s_{structural} = S_{structural}/N \qquad (37)$$

The deviation of observed structural entropy from the theoretical prediction for the reference random system can serve as a measure of regularity or complexity of genome architecture.

The structural entropy can also be used as a measure of fitting between empirical distribution of genome architecture elements and its chosen statistical approximation. Such correspondence may be assessed by the absolute difference

$$\Delta S = | S_{empirical} - <S_{structural}> | \qquad (38)$$

or in terms of $z$-ratio, $z = \Delta S / [<(\Delta S_{structural})^2>]^{1/2}$, obeying approximately to Gaussian statistics for the random deviations. In the particular example of the log-normal distribution the mean structural entropy is given by the second expression in Eq. (36). The dispersion $\sigma^2$ is related to the second moment of normalized length as $\ln <l^2> = \sigma^2$ (Eq. (28)). Thus, estimating parameter $\sigma^2$ through the logarithm of the second moment with empirical distribution and substituting it to the expression for the mean structural entropy corresponding to the log-normal distribution (Eq. (36)), the estimate for the difference (38) can be obtained. The similar estimates can also be obtained for the exponential or gamma-distributions. The smallest difference (38) would indicate the approximation with the best fitting empirical distribution.

**8. Discussion**

Our consideration proves that the interpretation of the observed regularities in genome architecture or genome tracks depends strongly on the chosen approximation and reference model. For example, the distribution of gene lengths in different organisms can equally well be fitted by both gamma- and log-normal distributions [9, 11]. As discussed in Sections 3 and 4, these distributions are associated with reciprocal processes of aggregation and fragmentation. Therefore, the genetic meaning of such correspondence and underlying evolution scenarios should be quite different in two cases. The structure of genes for eukaryotes is known to be broken [1], protein-coding exon regions are divided by the non-coding introns. The correlations between index α (Eq. (19)) and the mean number of exons would give evidence in favor of aggregation scenario. The complete study of eukaryotic genes ought to include additionally the separate and mutual analysis of exon and intron sets. The huge experimental material stored in the numerous databases cannot be used without proper theoretical background and standard reference models. The intricate relationships between chromatin folding and gene regulation as well as the possible evolution scenarios cannot be clarified without detailed statistical analysis of available data.

## Acknowledgements

The author is very grateful to Yu.V. Kravatsky and G.I. Kravatskaya for help in this work.

Table 1. Levels of chromosome folding in genomes of different organisms

A. Levels of minichromosome folding in virus SV-40 genome, $M$ = 5,224 bp

| Level | Unit | Characteristic length, bp | Reference |
|---|---|---|---|
| 1 | DNA wrapped around histone octamers and linker DNA | ~250 | [16] |
| 2 | Domains of DNA corresponding to the symmetry of the fifth order in icosahedral virion | ~$10^3$ | [16] |

B. Levels of chromosome folding in *E. coli* K-12 genome, $M$ = 4,639,674 bp

| Level | Unit | Characteristic length, bp | Reference |
|---|---|---|---|
| 1 | DNA wrapped around histone-like, nucleoid-associated proteins (NAP) | 100–150 | [17] |
| 2 | Clusters of DNA-NAP complexes (putative) | ~$10^3$ | |
| 3 | Looped microdomain | ~$10^4$ | [18] |
| 4 | 117 kb units of a plectoneme and of a solenoid containing 10–12 microdomains | ~$10^5$ | [18] |
| 5 | Six macrodomains (Ori, NS-right, Right, Ter, Left, and NS-left regions) | ~$10^6$ | [18] |

C. Levels of chromatin folding in human genome, $M \approx 3 \times 10^9$ bp

| Level | Unit | Characteristic length, bp | Reference |
|---|---|---|---|
| 1 | Nucleosome | 160–240 | [1] |
| 2 | Helix pitch of a 30-nm fiber comprising 6 nucleosomes | ~$10^3$ | [1] |
| 3 | Loop (300-nm coiled chromatin fiber) | ~$10^4$ | [1] |
| 4 | Set of 10–20 loops (700-nm coiled coil) | ~$10^5$ | [1] |
| 5 | Chromosome bands | ~$10^6$ | [1] |
| 6 | Long and short arms of a chromosome (euchromatin), telomere and centromere regions (heterochromatin regions) | ~$10^7$ | [1] |
| 7 | Chromosome | ~$10^8$ | [1] |

**Supplemental materials to Table 1**

A. Levels of minichromosome folding in virus SV-40 genome, $M = 5{,}224$ bp

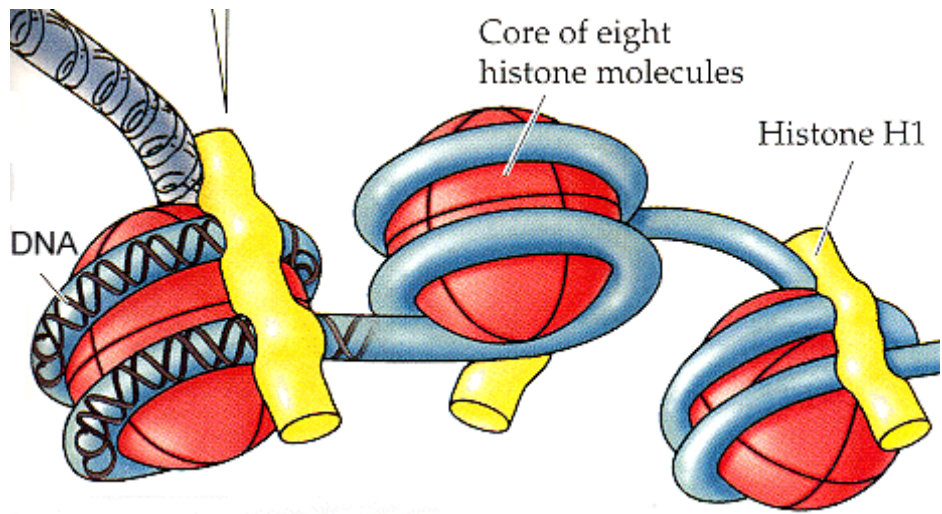

The first level of packing double-stranded DNA starts with wrapping DNA around histon-like proteins. The characteristic length of packing corresponds to the persistence length (about 150 bp). DNA wrapped around histon-like proteins is packed into capsid envelope (shown below).

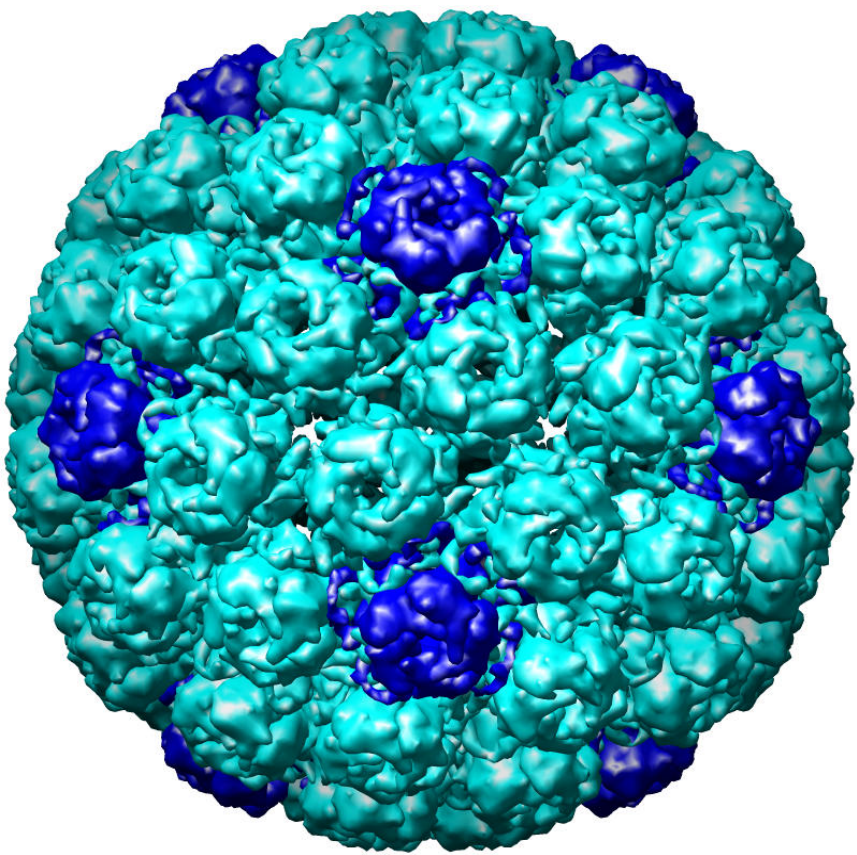

The schematic representation of capsid packing reveals icosahedral symmetry.

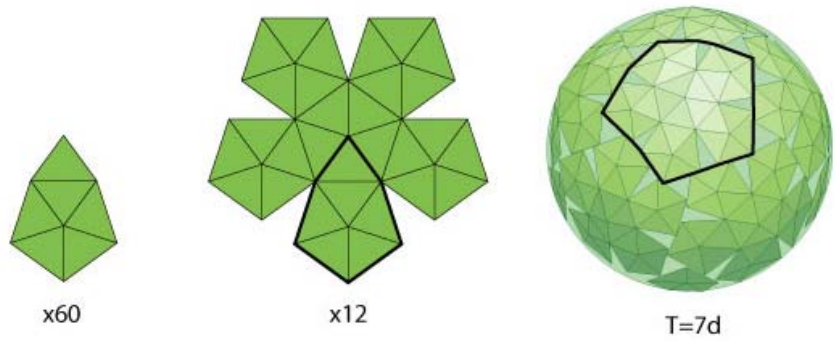

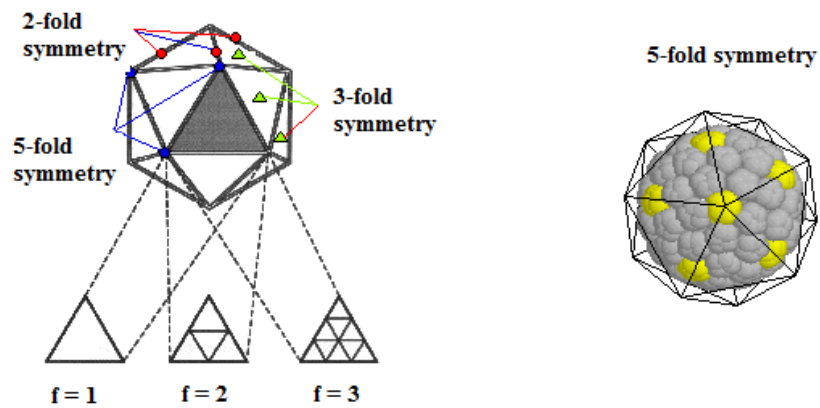

Domains of DNA corresponding to the symmetry of the fifth order in icosahedral virion can be attributed to the second level of folding. The symmetry of the fifth order in icosahedral capsid packing of viral genomes is associated with periods $M/5$ found in the viral genomic nucleotide sequences [20].

B. Levels of chromosome folding in *E. coli* K-12 genome, $M = 4{,}639{,}674$ bp

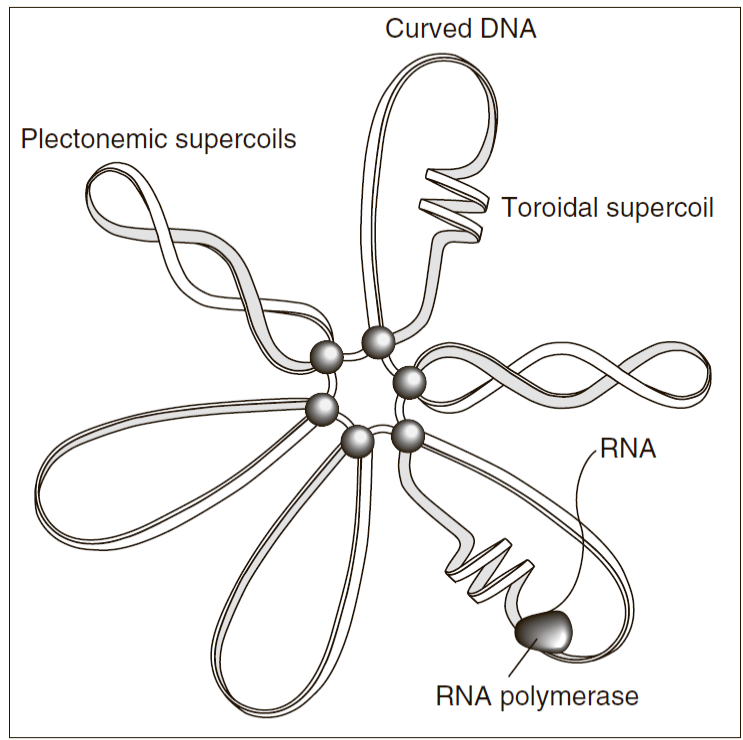

The first three levels of chromosome folding in *E. coli* are schematically shown in this figure (taken from Ref. [17]). The toroidal DNA is wrapped around histon-like, nucleoid-associated proteins (the first level). The second level is related to cooperative binding to histon-like proteins and subsequent packing of such strands (note that the characteristic length of this level is about the mean length of genes). The looped microdomains (their total number is about 400) correspond to the third level of folding. A part of DNA is free from histon-like proteins.

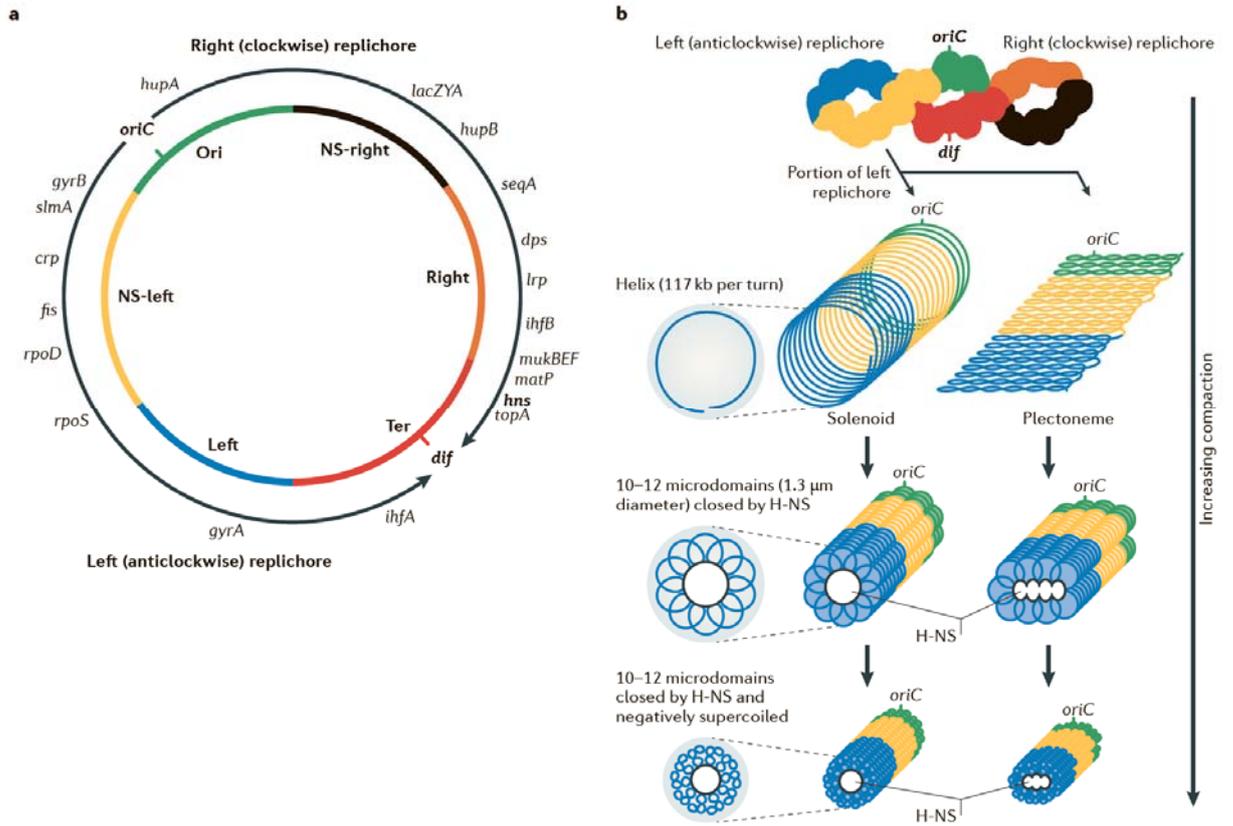

The levels from 3 to 5 (from bottom to top) are shown in this figure (taken from Ref. [18]).

C. Levels of chromatin folding in human genome, $M \approx 3\times10^9$ bp

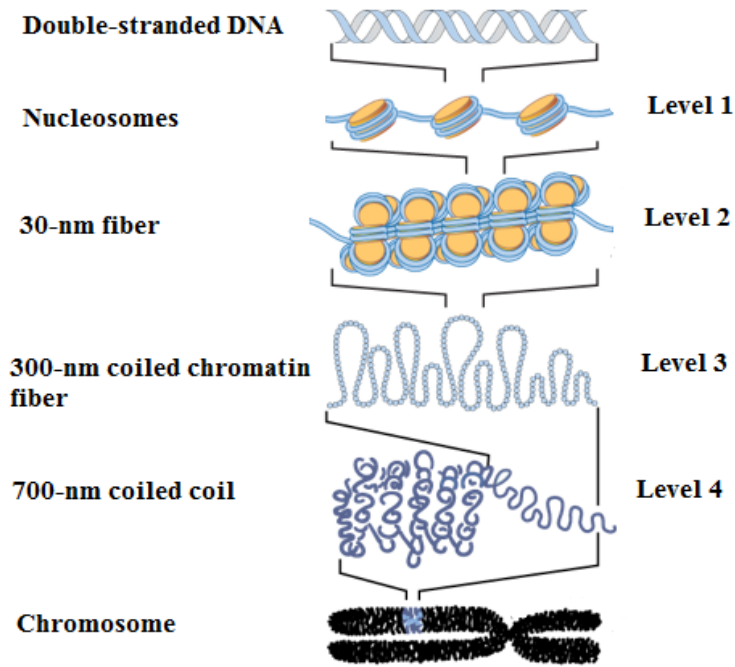

The levels 1–4 of chromatin folding in human genome are shown in this figure.

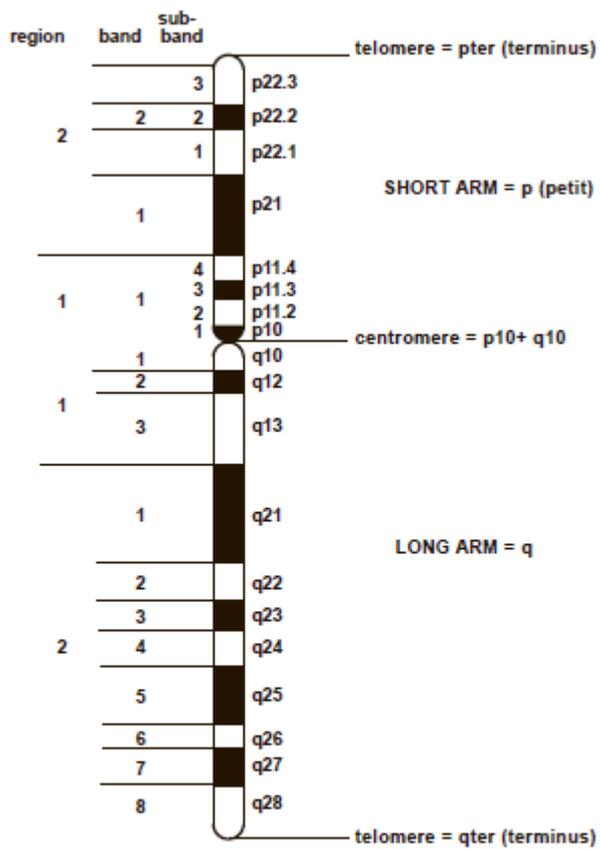

X Chromosome Band Pattern

The fifth level of chromatin folding in human genome corresponds to the chromosome bands displayed by dye staining.

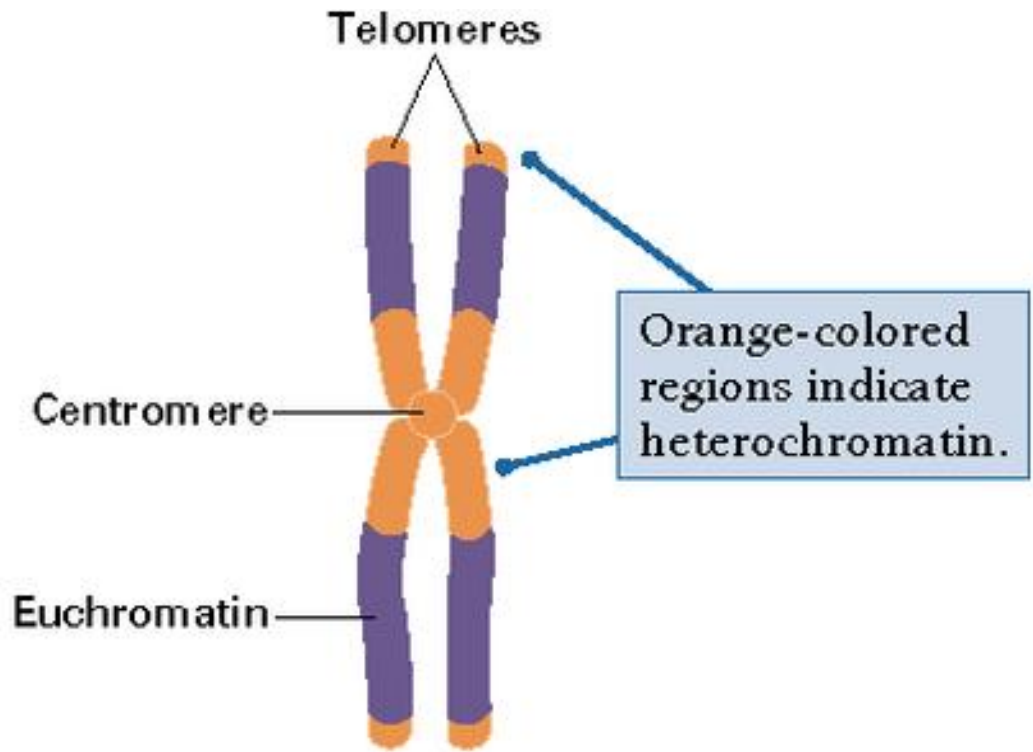

Euchromatin (less compact) and heterochromatin (more compact) regions correspond to the sixth level of folding, whereas the chromosome itself corresponds to the seventh level of folding.

The systematization of folding levels is yet absent. It is not clear whether is there one-to-one correspondence between all levels of folding throughout all chromosomal organisms independent of great variations in genome lengths.